\documentstyle[preprint,pre,aps]{revtex}

\newcounter{saveeqn}
\newcommand{\alpheqn}{\setcounter{saveeqn}{\value{equation}}
\stepcounter{saveeqn}\setcounter{equation}{0}
\renewcommand{\theequation}
      {\mbox{\arabic{saveeqn}-\alph{equation}}}}
\newcommand{\reseteqn}{\setcounter{equation}{\value{saveeqn}}
\renewcommand{\theequation}{{\arabic{equation}} } }
\title{Image restoration using the chiral Potts spin-glass}
\author{Domenico M. Carlucci} 
\address{Instituut voor Theoretische Fysica,\\
K. U. Leuven, B-3001 Leuven, Belgium }
\author{Jun-ichi Inoue} 
\address{ Department of Physics, Tokyo Institute of Technology,\\
Oh-okayama, Meguro-ku, Tokyo 152-8551, Japan}

\date{\today}
\begin{document}
\maketitle
\begin{abstract}
We report on the image 
reconstruction (IR) problem by making use of the random chiral 
$q$-state Potts model, whose Hamiltonian possesses the 
same gauge invariance as the usual Ising spin glass model.
We show that the pixel representation by means of 
the Potts variables is suitable for the gray-scale level image 
which can not be represented by the Ising model.  
We find that 
the IR quality is highly improved by the presence of a glassy term, besides 
the usual ferromagnetic term  under random external fields, 
as very recently pointed out by Nishimori and Wong.
We give the exact solution of the infinite range model
with $q=3$, the three gray-scale 
level case. 
In order to check our analytical result and the efficiency
of our model, 2D Monte Carlo simulations have been carried out on 
real-world pictures with three and eight gray-scale levels.
\end{abstract}
\mbox{}
\clearpage
\setcounter{page}{1} 
\section*{Introduction}
Recently, statistical mechanical approaches to the 
problems of information science attracted much attention of 
the researchers who are working in the field of information science. 
Among these,  
a particular interest has been given to the techniques by which 
one tries to reconstruct an image from its corrupted version, 
{\it e.g.} sent by a defective fax, a fickle {\it e}-mail {\it et cetera},  
since any data transmission through a channel is in principle
affected by some kind of noise.
In the mathematical engineering fields, 
the traditional way to obtain the optimal recovered image 
has been regarded as a sort of optimization problems. 
In this framework, one first constructs the energy (cost) function 
so that this function represents the distance between 
the original image and the recovered one as properly as possible;
then, one minimizes it using suitable heuristic methods like 
{\it simulated annealing} \cite{Kirkpatrick83}.
In fact, Geman and Geman \cite{Geman84} succeeded in 
constructing a method of image restoration 
using simulated annealing, and they discussed in detail
the properties of its convergence including the optimal annealing 
schedule.   

Successful results in this direction have been reached 
by means of the usual techniques of disordered 
spin systems, assuming that each spin is naturally associated to 
a pixel or bit. 
In the language of the disordered spin systems, 
the optimization problems we just mentioned are 
naturally translated into the search of the ground state a system 
possessing many local minima of order ${\exp}(N)$.   
In contrast, Marroquin {\it et al.}\cite{Marroquin87} 
found that the temperature of the system 
plays an important role for the image recovering process. 
From the statistical mechanical point of view, 
each recovered image can be regarded as the equilibrium state of 
a random spin systems.
Marroquin {\it et al.}\cite{Marroquin87} investigated the 
effect of the temperature on 
the quality of image restoration by computer simulation and 
found the optimality of finite temperature image restoration. 
Recently, this finite temperature effects on 
image restoration was checked in a more careful way 
by Pryce and Bruce\cite{PB}, 
although these works were restricted to numerical simulations.
In the context of the convolutional error-correcting codes, 
Ruj$\acute{\rm a}$n\cite{Rujan} proposed the finite temperature decoding 
in which we regard the sign of the local magnetization at a specific 
temperature (this temperature is well known as the   
{\it Nishimori temperature}\cite{Nishi} in the field of spin-glasses) 
as the correct bit. 
Recently Nishimori and Wong\cite{NW} pointed out that 
the optimal restoration of an image is also obtained at some specific 
temperature and 
showed that image restoration(IR) and error-correcting 
codes theory (ECC) can be treated within a single framework.
Indeed,  to the usual IR Hamiltonian, ferromagnetic and random 
field terms, they added a spin-glass term borrowed from 
the ECC theory\cite{Sourlas} used for {\it parity-check}.  
They could exactly solve the infinite range spin model and find the 
optimal values of the temperature and  the field (referred to as 
{\it hyper-parameters} from now on) at which the best retrieval quality 
is achieved.
However, their works are restricted to the case of 
Ising spin systems and in this sense they are able 
to restore black/white pictures. On the other hand,  
there remain many  open questions 
about the restoration of multi-color images or, somehow 
equivalently, gray toned images. 
This kind of problem 
has been also widely studied in the context of neural network 
with multi-state neurons, able to store and retrieve gray-scaled 
patterns (see \cite{Bolle} and references therein). 
For our purposes, we therefore map the set of the pixels 
onto $q$-state (chiral) Potts spins, with 
a ferromagnetic Hamiltonian in the presence 
of a random field (conventional IR) and, further on, 
a glass term (ECC-{\it like} term). The choice of the chiral 
Potts Hamiltonian is motivated by the fact that it exhibits 
the same gauge invariance as the Ising glass, although 
a work for the usual random Potts model is under consideration.
Here, we show that, as in the Ising case\cite{NW}, the presence 
of the glass term significantly 
increases the quality of the reconstructed image. 
We should mention that several remarkable studies 
about IR using the Potts model have been made by several authors. 
However, their works mostly depend on 
the computer simulations. 
In addition, their methods (mean field annealing\cite{Zhang}, 
cluster algorithm \cite{Tanaka96,Tanaka97}, 
{\it etc.}) are devoted to the restorations at zero temperature. 
Therefore, it seems that there exist a lot of open 
questions about IR using the Potts model, especially, 
about the performance of the finite temperature restoration.

In the next section, we will introduce our model 
within the image restoration theory and adopt 
the overlap as a 
measure of the restoration quality. 
In Sec. III, we will discuss the infinite range model 
and give the exact expression for the overlap as a function of 
the temperature and the external field, thus obtaining a relation between 
the temperature of source image and that of restoration temperature. 
We shall also see the improvement of the restoration quality by 
adding the glassy term. 
Finally, in Sec.  IV, guided by the  
infinite range results, we will give explicit and realistic 
examples of image reconstructions for three and eight gray-scale  
picture. 
\section{The model and IR formulation \label{model} } 
As already mentioned in Introduction, 
we choose to represent 
pixels of a gray-scaled image by means 
of $q$-component Potts spin variables.
The usual Potts Hamiltonian $H=-\sum \delta_{\sigma_i\sigma_j}$ 
admits a complex representation \cite{Stephen} 
by means of the following identity

\begin{equation}
  \delta_{\sigma_i\sigma_j}=\frac{1}{q}\sum_{r=0}^{q-1}
  (\sigma_i)^r  (\sigma_j)^{q-r}        
\end{equation}
where each spin takes on one of the $q$ roots of unity
\begin{eqnarray}
{\sigma}_{i}={\exp}
\left(
\frac{2 \pi i}{q}K_{i}
\right)\,\,\,\,\,\,\, \mbox{($K_{i}=0,{\cdots},q-1$)}.
\label{potts-spin}
\end{eqnarray}
From now on, we will use the notation $\{\xi\}$ for the original 
pixels and $\{\sigma\}$ for the variables of the 
recovering process. 
Let us now send the original image 
through a noise channel not only by the form of 
${\xi_i^r}$ itself but also by the following products 
${\xi}_{i}^r{\xi}_{j}^{r*}={\xi}_{i}^r{\xi}_{j}^{q-r}$.
Without loss of generality we raised the spins and 
their products  to some power $r$, since this corresponds only 
to a rotation in the complex circle. The reasons of this choice 
will be clear soon.  
For this expression, the output $(\{\tau^{(r)}\},\{J^{(r)}\})$ is 
stochastically determined  by the channel. For instance, in the 
case of a   Gaussian channel (GC) the   
output function $P_{\rm out}(\{J^{(r)}\},\{\tau^{(r)}\}|\{\xi\})$ 
is given by 
\begin{eqnarray}
P_{\rm out}(\{J^{(r)}\},\{\tau^{(r)}\}|\{\xi\}) & = & 
\frac{1}{(2\pi J)^{N_{B}/2}}\frac{1}{(2\pi \tau)^{N/2}} \nonumber \\
\mbox{} & \times & 
{\exp}\left[
-\frac{1}{2J^{2}}
\sum_{(ij)}\sum_{r=0}^{q-1}
(J_{ij}^{(r)}-J_{0}{\xi}_{i}^r{\xi}_{j}^{q-r})
(J_{ij}^{(r)*}-J_{0}{\xi}_{i}^{q-r}{\xi}_{j}^{r})
\right] \nonumber \\
\mbox{} & \times & 
{\exp}\left[
-\frac{1}{2\tau^{2}}
\sum_{i}\sum_{r=0}^{q-1}
({\tau}_{i}^{(r)}-{\tau}_{0}{\xi}_{i}^{r})
({\tau}_{i}^{(r)*}-{\tau}_{0}{\xi}_{i}^{q-r})
\right], 
\label{gc}
\end{eqnarray}
where $J_{ij}^{(r)}$ and $\tau_i^{(r)}$ are complex numbers which 
satisfy 

 \begin{equation} 
   \left(J_{ij}^{(r)}\right)^*\,=\,J_{ij}^{(q-r)}
   \,\,\,\,\,     
   \left(\tau_{i}^{(r)}\right)^*\,=\,\tau_{i}^{(q-r)}
   \label{condition}
 \end{equation} 
in order to insure the realness of the sums in (\ref{gc}).

Obviously, if a noise-free transmission  could be achieved, we would 
obtain
${\tau}_{i}^{(r)}={\xi}_{i}^r$ and 
$J_{ij}^{(r)}={\xi}_{i}^{q-r}{\xi}_{j}^{r}$.
The conditional probability 
$P(\{\sigma\}|\{J^{(r)}\},\{\tau^{(r)}\})$, 
which is probability that 
the source sequence is 
$\{\sigma\}$ provided that the outputs are $\{J\}$ and 
$\{\tau\}$, according to the Bayes theorem reads 
\begin{eqnarray}
P(\{\sigma\}|\{J^{(r)}\},\{\tau^{(r)}\})\,{\sim}\,
{\exp}\left(
\frac{{\beta}_{J}}{q}\sum_{(ij)}\sum_{r=1}^{q-1}
J_{ij}^{(r)}{\sigma}_{i}^{(r)}{\sigma}_{j}^{(q-r)}
+ \frac{h}{q} \sum_{i}\sum_{r=1}^{q-1}{\tau}_{i}^{(r)}{\sigma}_{j}^{(q-r)}
\right)P_{d}(\sigma)
\label{condprob}
\end{eqnarray}
where $P_{d}(\sigma)$ 
is a model of the prior distribution $P_{s}(\xi)$, that is, 
\begin{eqnarray}
P_{d}(\sigma)\,{\equiv}\, 
{\exp}\left(
\frac{\beta_{d}}{q}
\sum_{(ij)}\sum_{r=1}^{q-1}
{\sigma}_{i}^{(r)}{\sigma}_{j}^{(q-r)}
\right).
\label{modelP}
\end{eqnarray} 
Our choice of the above prior distribution (\ref{modelP})  
is due to the assumption that 
in real world, images should be locally smooth. 
From this point of view, the distribution (\ref{modelP}) 
is suitable because it 
gives high probability if the nearest neighboring sites take a same value.

For the Ising model, in order to get the restored pixels        
out of the average quantities, 
the pixel at site $i$ (to be denoted as $\Sigma_i$) 
is naturally taken as the sign of 
the local magnetization. 
This means that the restored pixel is chosen 
as $\Sigma=+1$ ($\Sigma=-1$) if the spin points upward 
(downward) on average at the equilibrium. 
For our model, instead, since the value of the local magnetization is not 
simply confined to the interval $[-1,1]$ but it runs all over the complex 
circle, we introduce the following generalized restored variable 
\begin{eqnarray}
\Sigma_{i}
\left(\langle \sigma_i \rangle \right)
\,=\,\exp
\left[
i\sum_{\alpha=0}^{q-1}
\frac{2\pi}{q}\,{\alpha}\,\Xi_{\alpha}({\theta}_{i})
\right]
\label{newsigma}
\end{eqnarray}
with 
\begin{eqnarray}
{\Xi}_{\alpha}(x)=
{\Theta}\left(
x-\frac{2\pi}{q}{\alpha}
+\frac{\pi}{q}
\right)
-
{\Theta}\left(
x-\frac{2\pi}{q}{\alpha}
-\frac{\pi}{q}
\right),
\label{defTheta}
\end{eqnarray}
$\Theta$ being the usual step function, and 
\begin{eqnarray}
{\theta}_{i}={\tan}^{-1}
\left(
\frac{\langle {\rm Re}\,[{\sigma}_{i}]\rangle}
{\langle {\rm Im}\,[{\sigma}_{i}]\rangle}
\right).
\label{angle}
\end{eqnarray}
In simpler words, $\Sigma_i$ is the closest spin on the circle to 
the value of the local magnetization  $\langle \sigma_i \rangle \equiv
\langle {\rm Re} \,[\sigma_i] \rangle +i \,\langle {\rm Im}\, [\sigma_i] 
\rangle$.
For $q=2$, it is straightforward to check that (\ref{newsigma}) 
reduces to a sign function up to a normalization constant. 
The quantities $\langle {\rm Re}\, [{\sigma}_{i}] \rangle$ 
and $\langle {\rm Im}\, [{\sigma}_{i}] \rangle$ 
are the average over the Boltzmann 
distribution ${\rm e}^{-{\cal H}_{\rm eff}}$ with the 
following 
effective Hamiltonian \cite{NS} 
\begin{eqnarray}
{\cal H}_{\rm eff} 
-\frac{\beta_{J}}{q}
\sum_{(ij)}\sum_{r=1}^{q-1}
J_{ij}^{(r)}({\sigma}_{i})^{r}({\sigma}_{j})^{q-r}
-\frac{\beta_{d}}{q}\sum_{(ij)}\sum_{r=1}^{q-1}
({\sigma}_{i})^{r}({\sigma}_{j})^{q-r}
-h\sum_{i}\sum_{r=1}^{q-1}
{\tau}_{i}^{(r)}
{\sigma}_{j}^{q-r}.
\label{effHam}
\end{eqnarray}
The condition (\ref{condition}) gives the above Hamiltonian 
the same spin gauge-symmetry as Ising spin glass, thus 
suppressing the spontaneous magnetization at low temperature 
which is present in the usual random Potts model.
For the restoration purposes, the random field term aligns the 
spins according to the corrupted picture, whereas the ferromagnetic 
term insures the smoothness, by suppressing the 
isolated pixels within one small cluster.
Therefore, a balance between 
${\beta}_{d}$ and $h$) will
helps us to reconstruct the original picture well. 
The first term, instead, has been  recently introduced in the problem of 
image restoration by Nishimori and Wong \cite{NW} and 
this term has been well known as the {\it parity check codes} 
in the field of error-correcting codes. 
Obviously, this term carries much more information about the 
original picture than the other two terms. 
Therefore, the performance of the image recovery is 
expected to be improved by this term. 
As a measure of the restoration quality, we shall adopt 
the following overlap $M$ 
\begin{eqnarray}
M & = 
& \left[
   \frac{1}{q}\sum_{r=0}^{q-1} 
    \xi_i^{q-r} \Sigma_{i}^r \right]_{ 
 \{  \xi, J, \tau \}          }
\,\equiv\,
\frac{1}{q}\sum_{r=0}^{q-1}
\sum_{\xi}\sum_{J}\sum_{\tau}
P_{\rm out}(\{J^{(r)}\},\{\tau^{(r)}\}|\{\xi\})P({\xi}) 
{\xi}_{i}^{q-r}  {\Sigma}_{i}^{r} 
 \label{overlap}
\end{eqnarray}
in which $(1/q)\sum_{r=0}^{q-1}\xi_{i}^{q-r}{\Sigma}_{i}^{r}$ 
at each single site gives $1$ if 
the original spin is in the same state as the restored one, 
and zero otherwise.
Here the dependence on the local magnetization is buried in the 
angle $\theta_i$, give by Eq. (\ref{angle}), 
and the sum over all the sites is understood.
The main goal of this paper is to maximize the overlap $M$ 
as a function of the temperatures ($\beta_J$ and $\beta_d$) 
and the external field $h$ (referred to as an estimate of 
the {\it hyper-parameters}). 
In the next section, we will start with an exactly solvable 
model, that is, an infinite range version of the Potts 
spin-glass.

\section{Mean field solution \label{meanfield}}
We will now investigate the performance of 
our model within the mean field approximation, {\it viz.} 
each spin is influenced by all the others. As the source image, 
we will consider a ferromagnetic state generated by a Boltzmann 
distribution at some finite temperature $T_s$.
For the sake of simplicity, we will restrict 
ourselves to the case of $q=3$, although the results 
can be generalized to any value of $q$.  
We thus assume that the original set of pixels $\{\xi\}$ is generated 
by a ferromagnetic 3-state Potts Hamiltonian with probability 
\begin{eqnarray}
P({\xi})=\frac{1}{{\cal Z}_{s}(\beta_{s})}
\,{\exp}\left[
\frac{\beta_{s}}{2N}
\sum_{i<j}
\left(
{\xi}_{i}{\xi}_{j}^{*}+{\xi}_{i}^{*}{\xi}_{j}
\right)
\right]
\end{eqnarray}
where ${\cal Z}_{s}(\beta_{s})$ 
is a normalization constant and $\beta_s$ is the inverse source 
temperature.
According to the conditional probability, 
the observable are computed as 
\begin{eqnarray}
\left[\langle f \rangle\right]_{ \{\xi, J, \tau \}}
\,=\,
\sum_{\xi}\sum_{J}\sum_{\tau}
P(\{J^{(r)}\},\{\tau^{(r)}\}|\{\xi\})P(\xi)
\frac{{\rm Tr}_{\sigma} f\,{\rm e}^{-{\cal H}_{\rm eff}}}{\cal Z}
\end{eqnarray}
with
\begin{eqnarray} 
{\cal Z}\,{\equiv}\,{\rm Tr}_\sigma{\exp}\left(-{\cal H}_{\rm eff}\right).
\end{eqnarray}
It is rather straightforward to average out the disorder by means 
of the well-known replica trick \cite{SK} and, assuming 
replica symmetry ansatz and isotropy (no dependence on $r$), 
the saddle point equations for the order parameters are given by 
\begin{eqnarray}
[\langle{\sigma}_{i}^{r}\rangle]{\equiv}m & = & 
\frac{1}{{\cal Z}_{s}}
\sum_{\xi}{\rm e}^{{\beta}_{s}(m_{s}^{(1)}{\rm Re}[\xi]
+m_{s}^{(2)}{\rm Im}[\xi])}\int\frac{du}{\sqrt{\pi}}\frac{dv}{\sqrt{\pi}}
{\rm e}^{-u^{2}-v^{2}}\frac{Z_{\rm cos}(\xi)}{Z(\xi)} \label{sigma}  \\
\mbox{}[{\rm Re}[{\xi}_{i}]\langle{\sigma}_{i}^{r}\rangle]{\equiv}t_{1} & = & 
\frac{1}{{\cal Z}_{s}}
\sum_{\xi}{\rm Re}[\xi]{\rm e}^{{\beta}_{s}(m_{s}^{(1)}{\rm Re}[\xi]
+m_{s}^{(2)}{\rm Im}[\xi])}\int\frac{du}{\sqrt{\pi}}\frac{dv}{\sqrt{\pi}}
{\rm e}^{-u^{2}-v^{2}}\frac{Z_{\rm cos}(\xi)}{Z(\xi)} \\
\mbox{}[{\rm Im}[{\xi}_{i}]\langle{\sigma}_{i}^{r}\rangle]{\equiv}t_{2} & = & 
\frac{1}{{\cal Z}_{s}}
\sum_{\xi}{\rm Im}[\xi]{\rm e}^{{\beta}_{s}(m_{s}^{(1)}{\rm Re}[\xi]
+m_{s}^{(2)}{\rm Im}[\xi])}\int\frac{du}{\sqrt{\pi}}\frac{dv}{\sqrt{\pi}}
{\rm e}^{-u^{2}-v^{2}}\frac{Z_{\rm cos}(\xi)}{Z(\xi)} \\
\mbox{}[\langle{\sigma}_{i}^{r}\rangle \langle {\sigma}_{j}^{q-r} \rangle]{\equiv}Q & = & 
\frac{1}{{\cal Z}_{s}}
\sum_{\xi}{\rm e}^{{\beta}_{s}(m_{s}^{(1)}{\rm Re}[\xi]
+m_{s}^{(2)}{\rm Im}[\xi])} \nonumber \\
\mbox{} & \times & \int\frac{du}{\sqrt{\pi}}\frac{dv}{\sqrt{\pi}}
{\rm e}^{-u^{2}-v^{2}}
\frac{1}{Z^{2}(\xi)}[Z_{\rm cos}^{2}(\xi)+Z_{\rm sin}^{2}(\xi)]
\label{Q} 
\end{eqnarray} 
Here $m_{s}^{(1)}$ and  $m_{s}^{(2)}$ are simply  
the real and imaginary components of the source magnetization,  
{\it viz} the usual non-random Potts model \cite{Stephen} 
mean field equations 
\begin{eqnarray}
[{\rm Re}[{\xi}_{i}]]\,{\equiv}\, m_{s}^{(1)} & = & 
\frac{1}{{\cal Z}_{s}}
\left(
{\rm e}^{\beta_{s}m_{s}^{(1)}}
-{\rm e}^{-\frac{1}{2}\beta_{s}m_{s}^{(1)}}{\cosh}
\left[
(\sqrt{3}/2)
{\beta}_{s}m_{s}^{(2)}
\right] 
\right) \\
\mbox{}[ {\rm Im} [\xi_{i}] ] \,{\equiv}\, m_{s}^{(2)} & = & 
\frac{1}{        {\cal Z}_{s}}
\sqrt{3}\,{\rm e}^{-\frac{1}{2}\beta_{s}m_{s}^{(1)}}
{\sinh}\left[
(\sqrt{3}/2)
{\beta}_{s}m_{s}^{(2)}
\right]
\end{eqnarray}
and
\alpheqn
\begin{eqnarray}
{\cal Z}_{s} & \,=\, & {\rm e}^{{\beta}_{s}m_{s}^{(1)}}
+2{\rm e}^{-\frac{1}{2}{\beta}_{s}m_{s}^{(1)}}
{\cosh}\left[
(\sqrt{3}/{2})\,  {\beta}_{s}m_{s}^{(2)} 
\right] \\
Z_{\rm cos}(\xi) & \,=\, & {\rm e}^{U(\xi)}-
{\rm e}^{-\frac{1}{2}U(\xi)}{\cosh}V \\
Z_{\rm sin}(\xi) & \,=\, & 
\sqrt{3}\,{\rm e}^{-\frac{1}{2}U(\xi)}
{\sinh}V \\
Z(\xi)&  \,=\, & {\rm e}^{U(\xi)}
+2{\rm e}^{-\frac{1}{2}U(\xi)}{\cosh}V
\end{eqnarray} 
\reseteqn
with
\alpheqn
\begin{eqnarray}
U & = & u\left[
\frac{\beta_{J}^{2}J^{2}}{q^{2}}Q
+{\tau}^{2}h^{2}
\right]^{\frac{1}{2}}
+\frac{\beta_{d}}{q}m
+\frac{\beta_{J}J_{0}}{q}
\left[
t_{1}{\rm Re}[\xi]
+t_{2}{\rm Im}[\xi]
\right]
+{\tau}_{0}h{\rm Re}[\xi] \\
V & = & \frac{\sqrt{3}}{2}\frac{\beta_{J}J}{q}
Q^{\frac{1}{2}}v.
\end{eqnarray}
\reseteqn
Finally, the overlap $M$ is expressed as the following weighted average 
\begin{equation}
M \,=\,  
\frac{1}{{\cal Z}_{s}}
\sum_{\xi}{\rm e}^{{\beta}_{s}(m_{s}^{(1)}{\rm Re}[\xi]
+m_{s}^{(2)}{\rm Im}[\xi])} 
 \int_{{\cal S}(\xi)}\frac{du}{\sqrt{\pi}}\frac{dv}{\sqrt{\pi}}
{\rm e}^{-u^{2}-v^{2}}
\label{OVERLAP}
\end{equation}
receiving contributions from the following $q=3$ regions in the 
complex circle

\alpheqn

\begin{eqnarray}
{\cal S}(1) & = & 
\left\{
u,v\,{\Biggr |}\,
-\frac{\pi}{3}\,{\leq}\,{\tan}^{-1}\frac{Z_{\rm sin}}{Z_{\rm cos}}\,{\leq}\,
\frac{\pi}{3} \,\bigcap \, Z_{\rm cos}>0
\right\} \\
{\cal S}({\rm e}^{\frac{2\pi i}{3}}) & = & 
{\Biggr \{}
u,v\,{\Biggr |}\,
\left(
\frac{\pi}{3}\,{\leq}\,{\tan}^{-1}\frac{Z_{\rm sin}}{Z_{\rm cos}}\,
{\leq}\,
\frac{\pi}{2}  \,\bigcap \, Z_{\rm cos} \,{\geq}\, 
0, Z_{\rm sin} > 0
\right) \nonumber \\
\mbox{} & \mbox{} & \bigcup 
\left(-\frac{\pi}{2}\,{\leq}\,{\tan}^{-1}
\frac{Z_{\rm sin}}{Z_{\rm cos}}\,{\leq}\,
0 \,\bigcap\, Z_{\rm cos}\,{\leq}\, 0, Z_{\rm sin} \,{\geq}\, 0
\right)
{\Biggr \}} \\
{\cal S}({\rm e}^{\frac{4\pi i}{3}}) & = &  
{\Biggr \{}
u,v \,{\Biggr |}\,
\left( 
\frac{\pi}{2}\,{\leq}\,{\tan}^{-1}
\frac{Z_{\rm sin}}{Z_{\rm cos}}\,{\leq}\,
-\frac{\pi}{3}\,\bigcap\, 
Z_{\rm cos} >  0, Z_{\rm sin} \,{\geq}\, 0
\right)  \nonumber \\
\mbox{} & \mbox{} & \, \bigcup\, 
\left(0\,{\leq}\,{\tan}^{-1}\frac{Z_{\rm sin}}{Z_{\rm cos}}\,{\leq}\,
\frac{\pi}{2}\, \bigcap\,Z_{\rm cos} <  0, Z_{\rm sin} \,{\leq}\, 0
\right)
{\Biggr \}}.
\end{eqnarray}
\reseteqn

We first assume that the exchange term is absent (${\beta}_{J}=0$) \cite{Nishi83}, that 
is no redundancy is fed into the channel. 
In this case, the saddle point equations (\ref{sigma}-\ref{Q}) are drastically 
simplified and the overlap (\ref{OVERLAP}) simply reads 
\begin{equation}
M =  
\frac{{\rm e}^{\beta_{s}m_{s}}}
{{\cal Z}_{s}}
{\rm Erf}\left[
-\sqrt{2}\frac{{\beta}_{d}m+{\tau}_{0}h}{\tau h}
\right]
+
\frac{{\rm e}^{-\frac{1}{2}}}
{{\cal Z}_{s}}
\left\{
1-{\rm Erf}\left[
-\sqrt{2}\frac{{\beta}_{d}m-{\tau}_{0}h}{\tau h}
\right]
\right\}
\end{equation}
with the magnetization given by 
\[
m =  
\frac{{\rm e}^{{\beta}_{s}m_{s}}}
{{\cal Z}_{s}}
\int\frac{du}{\sqrt{\pi}} {\rm e}^{-u^{2}}
\frac{1-{\exp}[-\frac{3}{2}(u\tau h +{\beta}_{d}m/q+{\tau}_{0}h)]}
{1+2{\exp}[-\frac{3}{2}(u\tau h +{\beta}_{d}m/q+{\tau}_{0}h)]}
\]

\begin{equation}
\mbox{} +  
\frac{2{\rm e}^{-\frac{1}{2}{\beta}_{s}m_{s}}}
{{\cal Z}_{s}}
\int\frac{du}{\sqrt{\pi}}
{\rm e}^{-u^{2}} 
\frac{1-{\exp}[-\frac{3}{2}(u\tau h +{\beta}_{d}m/q-{\tau}_{0}h/2)]}
{1+2{\exp}[-\frac{3}{2}(u\tau h +{\beta}_{d}m/q-{\tau}_{0}h/2)]}
\end{equation}
where we defined ${\rm Erf}(x)\,{\equiv}\,\int_{x}^{\infty}{\rm e}^{-x^{2}}dx
/\sqrt{\pi}$.

The problem is thus reduced to a one-dimensional model, 
corresponding to an Ising model in which 
the length of spin turns out to be  $(+1,-1/2)$ instead of 
$(+1,-1)$. This is not surprising if one thinks that
 the fluctuations along the imaginary axis are governed only by the glassy
term, meanwhile the magnetic field acts along the real 
direction. In Fig. \ref{q3MTd_inf}, we plotted the 
overlap $M$ as a function of $T_{d}$ 
for the some values of $h$. It is straightforward to check 
that the maximum value of the overlap $M_{\rm max}$ does not 
depend on magnetic field $h$, since at the stationary point 
$\left(\partial M/\partial \beta_d=0\right)$ $m$ is proportional 
to the magnetic field 

\begin{eqnarray}
\frac{1}{2}
{\beta}_{s}m_{s}=
\frac{1}{3}m{\beta}_{d}
\frac{\tau_{0}}{\tau^{2}h} 
+\frac{1}{4}
\frac{{\tau}_{0}^{2}}{\tau^{2}}.
\label{relation}
\end{eqnarray} 
This feature holds also for the Ising case, although the 
stationary equation (\ref{relation}) is simpler 
\begin{eqnarray}
{\beta}_{s}m_{s}=m{\beta}_{d}\frac{{\tau}_{0}}{{\tau}^{2}h}. 
\end{eqnarray} 
Expression (\ref{relation}) is thought to be valid only for the infinite 
range model, as confirmed in the next section by numerical results 
in $d=2$. 

Now we set the decoding temperature at the optimal value, 
that is $M(T_d^{\rm opt})\equiv M_{\rm max}$ and we switch 
the exchange interaction ($\beta_J \neq 0$) as depicted in 
Fig. \ref{q3Mbeta_inf}. 
We notice that also a small amount of redundancy 
highly improves the value of the overlap $M_{\rm max}$ which quickly 
increases and slowly decreases, after the peak, meanwhile 
the exchange term becomes dominant on the ferromagnetic one.
\section{Monte Carlo simulations for real world picture \label{Montecarlo}}
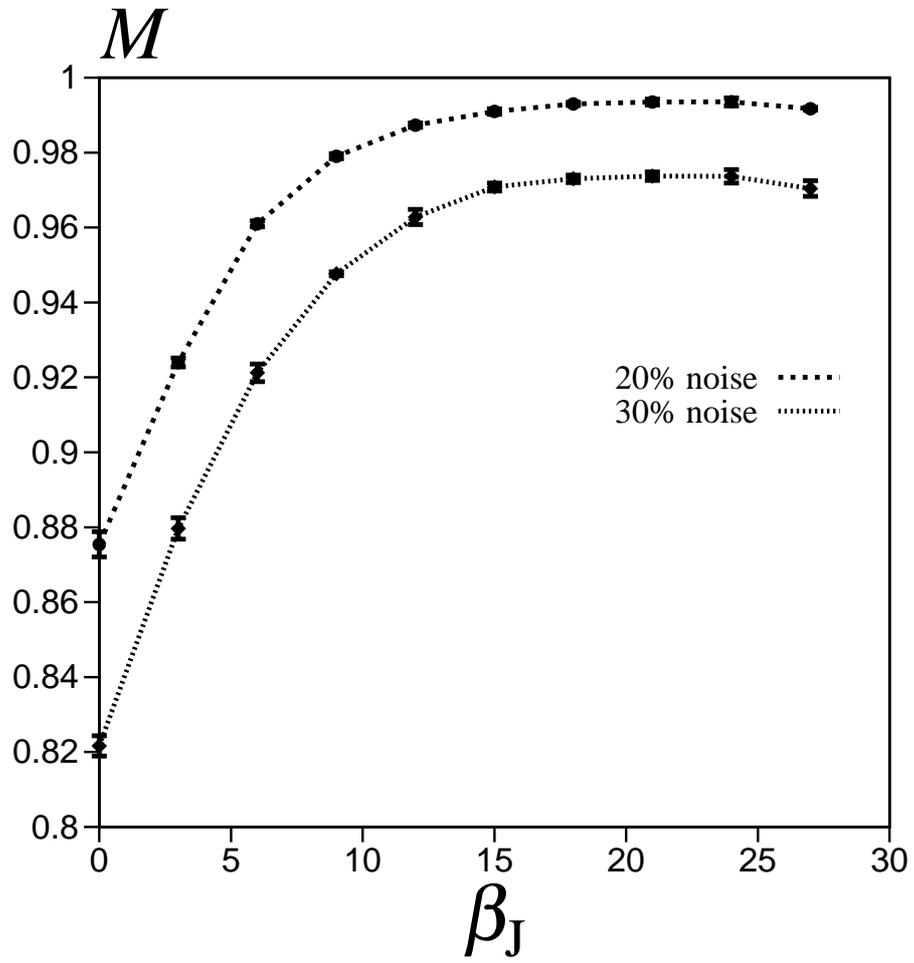
\begin{figure}
\caption{ Overlap as a function of ${\beta}_{J}$.  We set the 
                   parameters $(H,T_{d})=(0.6,0.1)$ which gives the
		   maximum in the absence of the exchange term. }
 \label{q8Mbeta}
\end{figure}

Although for the mere restoration aims it is not wise to
smoothen two points far away from each other, 
we shall see that the infinite range model provides a useful guide 
for the more interesting case of real-world
pictures, since the results remain qualitatively similar.
We thus carried out Monte Carlo simulations for 
realistic pictures with short-range effective Hamiltonian. 
In this case, the ferromagnetic term will 
be concerned  only with the points within the range of interaction and two 
points far away will not influence each other. 
It would be extremely interesting to study the restoration quality 
as a function of the interaction radius, but this goes beyond the aim
of the present work and we limit ourselves to a first nearest-neighbors 
interaction Hamiltonian.
Therefore let us consider a simple $q=3$ gray scale level picture 
(upper left of Fig. \ref{q3noise15}), where each pixels 
has been randomly flipped to another value with some probability, say
$p=0.15$ (upper right of Fig. \ref{q3noise15} ).
The curves shown in Fig.\ref{q3MTd} are the result of the 
restoration process without the glassy term, that is  
$\beta_J=0$, at different values of the ratio $H=h/\beta_d$. 
Here the maximum value of the overlap is achieved around 
$H_{\rm max}{\equiv}h/{\beta}_{d}\,{\sim}\,0.6$ and $T_{d}\,{\sim}\,0.2$ and
the corresponding restored image is drawn in the lower left of 
Fig. \ref{q3noise15}.
Adding the glassy term at $H_{\rm max}$ fixed improves drastically the 
value of the overlap and the quality of the restored 
image (lower right of Fig. \ref{q3noise15}), drawn at the peak of the 
Fig. \ref{q3Mbeta}. 
The same procedure is repeated in the presence of higher noise,
$p=0.30$, at the same $H_{\rm max}$ and $\beta_{J;{\rm max}}$ and the 
results of the restoration are shown in  Figs. \ref{q3noise30}. 
Finally, we applied the same algorithm to an eight gray-scale level 
picture with $20\%$ and $30\%$ of noise, upper images in Figs. 
\ref{q8noise20} and \ref{q8noise30}.
The results without exchange term are shown in Fig. \ref{q8MTd}. Once 
again we find a maximum for some values of $T_d$ and $H$ 
and the corresponding restored images are in Fig. \ref{q8Mbeta}.

\section{Conclusions}
In this paper, we investigated 
the possibility of gray-scaled image restoration 
using chiral random Potts model.
We exactly solved the infinite ranged version thus 
deriving the explicit expression of the   
overlap as a function of the estimates of the hyper-parameters 
$h$, ${\beta}_{d}$ and ${\beta}_{J}$. 
In the absence of the glassy term, 
we obtained the exact relation between 
the restoration temperature ${\beta}_{d}$ and 
the source temperature ${\beta}_{s}$ 
which gives the maximum value of the overlap.
This seems a highly non-trivial result because 
it is natural for us to 
assume that the best recovery of the image should 
be achieved for ${\beta}_{d}={\beta}_{s}$, as it turns 
out to be true for the Ising case\cite{NW}.
The Monte Carlo results on real pictures 
confirmed the expected high improvement due to 
the presence of the redundancy, {\it i.e.} 
the glassy term. However, so far, in our 
prescription to recover a corrupted image at the best restoration 
values, one is supposed to know the original data. 
In other words, the receiver has to meet the sender 
at least once to find the optimal restoration values.
Only after that, the other receivers 
will be able to get an optimal restoration for the same image,
provided that the channels remain, at least qualitatively, 
unchanged.
In this sense, it would be extremely useful to provide some 
{\it a priori} criteria (the receiver will not be supposed 
to meet the sender) for the optimum values of the 
hyper-parameters, once that some intrinsic characteristics 
({\it e.g.} temperature) of the original image are known.
Therefore, in order to check if the relation (\ref{relation}) 
still holds down to two dimensions, we restored $q=3$ ferromagnetic 
snapshots generated at some known temperature.
However, so far  we have not yet obtained reliable results and 
detailed investigations in this direction will be reported in
a forth coming paper. 
%

\section*{acknowledgement}

We thank Prof. Hidetoshi Nishimori for useful discussions and 
showing us his paper prior to publication. 
We also thank Prof. Kazuyuki Tanaka for useful advice and fruitful discussions.

The authors were supported by JSPS-Royal Society/British Council 
Anglo-Japanese Scientific Cooperation Programme. 
One of the authors (D.M.C.) started this work under 
JSPS grant No.P96215.

\clearpage


\begin{figure}
\caption{Overlap $M$ as a function of  $T_{d}$ for different values of $h$. 
The maximum value $M_{\rm max}$ does not depend on $h$.}
\label{q3MTd_inf}
\end{figure}

\begin{figure}
\caption{Overlap $M_{\rm max}$ as a function of the exchange temperature 
$\beta_J$ for several values of $J_0$. The overlap improves even for 
small values of the exchange term.}
\label{q3Mbeta_inf}
\end{figure}

\begin{figure}
\caption{ Upper left: original three gray-scale levels image. 
                    Upper right: 15\% of noise. Lower left: restoration 
                    without exchange term. Lower right:
                    restoration with exchange term. }
\label{q3noise15}
\end{figure}

\begin{figure}
\caption{  Upper left: original three gray-scale levels image. 
                    Upper right: 30\% of noise. Lower left: restoration 
                    without exchange term. Lower right:
                    restoration with exchange term.   }

\label{q3noise30}
\end{figure}

\begin{figure}
\caption{  Overlap $M$ as a function of the decoding temperature 
                    $T_{d}$  at $\beta_J=0$ (left). The system size is 
                   $64{\times}72$ and each line is averaged over
		   different $4$ samples. }
 \label{q3MTd}
\end{figure}

\begin{figure}
\caption{ Overlap as a function of ${\beta}_{J}$.  We set the 
                   parameters $(H,T_{d})=(0.6,0.2)$ which gives the
		   maximum in the absence of the exchange term.}
 \label{q3Mbeta}
\end{figure}

\begin{figure}
\caption{  Upper left: original eight gray-scale levels image. 
                    Upper right: 20\% of noise. Lower left: restoration 
                    without exchange term. Lower right:
                    restoration with exchange term.   }

 \label{q8noise20}
\end{figure}

\begin{figure}
\caption{  Upper left: original eight gray-scale levels image. 
                    Upper right: 30\% of noise. Lower left: restoration 
                    without exchange term. Lower right:
                    restoration with exchange term.   }

 \label{q8noise30}
\end{figure}

\begin{figure}
\caption{  Overlap $M$ as a function of the decoding temperature 
                    $T_{d}$  at $\beta_J=0$ (left). The system size is 
                   $93{\times}100$ and each line is averaged over
		   different $4$ samples.}
 \label{q8MTd}
\end{figure}

\begin{thebibliography}{99}

\bibitem{Kirkpatrick83}
S. Kirkpatrick, C. D. Gelatt and M. P. Vecchi, 
Science {\bf 220}, 671 (1983). 

\bibitem{Geman84}
S. Geman and D. Geman, IEEE Trans. on  Pattern  Anal.  Mach.  Intel. {\bf 6}, 721 (1984).

\bibitem{Marroquin87}
J. Marroquin, S. Mitter and T. Poggio, 
J. American Stat. Assoc. {\bf 82}, 76 (1987).

\bibitem{PB}
J. M. Pryce and A. D. Bruce, 
J. Phys. A: Math. Gen. {\bf 28}, 511 (1995). 


\bibitem{Rujan}
P. Ruj$\acute{\rm a}$n, Phys. Rev. Lett. {\bf 70}, 2968 (1993).

\bibitem{Nishi}
H. Nishimori, Prog. Theor. Phys.  {\bf 66}, 1169 (1981).

\bibitem{NW}
H. Nishimori and K. Y. M. Wong, {\it unpublished}. 

\bibitem{Sourlas}
N. Sourlas,  Nature {\bf 339},  693(1989).


\bibitem{Bolle} 
D. Boll\'e, B. Vinck and V. A. Zagrebnov, J. Stat. Phys. {\bf 70}, 1099 (1993).


\bibitem{Zhang}
J. Zhang, IEEE Trans. Signal Process. {\bf 40}, 2570 (1992).

\bibitem{Tanaka96}
K. Tanaka and T. Morita,`` Theory and Applications of the 
Cluster Varidation and Path Probability Methods'', 
Edited by J. L. Mor$\acute{\rm a}$n-L$\acute{\rm o}$pez and 
J. M. Sanches, 
Plenum Press, New York 1996. 


\bibitem{Tanaka97}
T. Morita and K. Tanaka, Pattern Recognition Letters {\bf 18}, 
1479 (1997). 


\bibitem{Stephen}
L. Mittag and M. J. Stephen, J. Phys. A {\bf 7}, L109 (1974).


\bibitem{NS}
H. Nishimori and M. J. Stephen, Phys. Rev. B  {\bf 27}, 5644 (1983).

\bibitem{Nishi83}
H. Nishimori, Phys. Rev. B {\bf 28}, 4011 (1983). 

\bibitem{SK}
D. Sherrington and S. Kirkpatrick, Phys. Rev. Lett. {\bf 35}, 1792 (1975).



\end{thebibliography}
\end{document}